\documentclass[twocolumn,showpacs,preprintnumbers]{revtex4}
\usepackage{amsmath}

\usepackage{amssymb}
\usepackage{mathrsfs}
\usepackage{bbm}
\usepackage{graphicx}
\usepackage{dcolumn}
\usepackage{bm}
\begin{document}

\title{ First-principles study of the incorporation and diffusion of helium
in cubic zirconia }
\author{Peng Zhang$^{1,2}$}
\author{Yong Lu$^{2,3}$}
\author{Chao-Hui He$^{1}$}
\author{Ping Zhang$^{2,4}$}
\thanks{Author to whom correspondence should be addressed. E-mail:
zhang\_ping@iapcm.ac.cn}
\affiliation{$^{1}$Department of Nuclear Science and Technology, Xi'an Jiaotong
University, Xi'an 710049, People's Republic of China\\
$^{2}$LCP, Institute of Applied Physics and Computational Mathematics,
Beijing 100088, People's Republic of China\\
$^{3}$Department of Physics, Beijing Normal University, Beijing
100875, People's Republic of China\\
$^{4}$Center for Applied Physics and Technology, Peking University,
Beijing 100871, People's Republic of China}

\begin{abstract}
The incorporation and diffusion of helium (He) with and without
intrinsic vacancy defects in cubic ZrO$_{2}$ are investigated
through first-principles total-energy calculations, in which the
projector-augmented-wave (PAW) method with the generalized gradient
approximation (GGA) is used. The calculated formation energies of
intrinsic point defects indicate that cubic ZrO$_{2}$ has a tolerant
resistance to radiation damage. The incorporation energy of He
impurity shows that it is preferable to occupy the Zr vacancy at
first, whereas the solution energy suggests that He would be
accommodated in the interstitial site at thermodynamic equilibrium
concentration. By calculating the He migration energies
corresponding to both interstitial and vacancy assisted mechanisms,
we suggest that it is most likely for He to diffuse by hopping
through a single vacancy. Remarkably, our calculated
vacancy-assisted diffusion energy of He is consistent well with the
experimental measurement.
\end{abstract}

\maketitle

\section{INTRODUCTION}

Zirconia (ZrO$_{2}$) is a highly attractive material of great
interests both scientifically and in terms of its technological
applications due to its
excellent mechanical, thermal, chemical, and dielectric properties \cite%
{Fadda1}. For example, ZrO$_{2}$ is employed as superplastic structural
ceramic that demonstrates superb strength and fracture toughness \cite%
{Garvie}, and is also used as one of the best refractive engineering
materials for thermal barrier coating on aeronautical and land-based gas
turbine blades \cite{Fevre}. Additionally, the cation-doped ZrO$_{2}$ can
also be used in solid oxide fuel cells and oxygen sensors \cite{Smart}.
Furthermore, as one of the most radiation resistant ceramic materials, ZrO$%
_{2}$ is applied as a promising candidate host phase for the inert matrix
nuclear fuel (IMF) in order to immobilize \cite{Curti} and burn up \cite%
{Gong} the plutonium from dismantled nuclear weapons and other high level
nuclear waste (HLW) caused by the actinides (e.g. $^{241}$Am). Consequently,
great attentions are needed to focus on the basic scientific research of ZrO$%
_{2}$.

ZrO$_{2}$ exists at least five crystalline structures with different
symmetries. The monoclinic \emph{P2$_{1}$/c} polymorph is the only
one found at ambient condition whereas the tetragonal
\emph{P4$_{2}$/nmc} and the cubic \emph{Fm$\bar{3}$m} phases are
stable above 1400 and 2600 K and the other two orthorhombic
\emph{Pbca} and \emph{Pnma} phases are stable above 3 and 20 GPa,
respectively \cite{Ohtaka1,Haines1,Haines2,Arashi,Ohtaka2}. As these
five ZrO$_{2}$ structures are centrosymmetric, they are all nonpolar
and nonpiezoelectric. Presently, ZrO$_{2}$ stabilized in the cubic
phase with e.g. Y$_{2}$O$_{3}$ is considered as a suitable
refractory host for actinide confinement used for transmutation or
high-temperature reactor projects. Large amounts of experimental
measurements have been carried out to determine the stability of
this material in high radiation environment. One early investigation
on phase stability of ZrO$_{2}$ under fission fragment damage was
reported by Wittels and Sherrill \cite{Wittels}. Then the evolution
of radiation induced damages in ZrO$_{2}$ \cite{Sickafus} was
systematically investigated using Rutherford backscattering
spectrometry and ion channeling (RBS/C), along with X-ray
diffraction and transmission electron microscopy (TEM) under Xe and
I ion irradiation. No amorphization of the pure ZrO$_{2}$ was
observed. But Meldrum \textit{et al}. \cite{Meldrum} found that
nanocrystalline ZrO$_{2}$ could be amorphized under Ne and Xe ion
irradiation. Damage evolution, surface morphology and characteristic
structural changes in He-implanted cubic ZrO$_{2}$ were extensively
investigated in experiments \cite{Kuri1,Kuri2,Kuri3,Velisa}. Fluence
dependence and thermal stability of defects in He-implanted cubic
ZrO$_{2}$ was determined by slow positron implantation spectroscopy
\cite{Saude}. Moreover, the He damage and He effusion in fully
stabilized ZrO$_{2}$ \cite{Damen} was measured with neutron depth
profiling (NDP) for different annealing temperatures and analyzed by
electron microscopy during the various annealing stages concluding
that the diffusion of He is probably caused by vacancy assisted
interstitial diffusion. Besides, He migration in monoclinic and
cubic yttria-stabilized ZrO$_{2}$ was also studied from
non-destructive $^{3}$He depth profiling using the resonant
$^{3}$He(d,p)$^{4}$He nuclear reaction \cite{Costantini}.

From theoretical point of view, \emph{ab initio} methods have been
employed to study the structural, electronic, mechanical,
thermodynamical, and vibrational
properties of ZrO$_{2}$ \cite%
{Fadda1,Jansen,Orlando,French,Jomard,Lowther,Dewhurst,Dash,Jaffe,Sternik,Kuwabara,Fadda2}%
. Very recently, the energetics of intrinsic point defects and the
influence of Al doping in ZrO$_{2}$ \cite{Foster, Zheng, Arhammar}
have been studied to identify the dominant defects under different
oxygen chemical potentials and Fermi levels. However, up to date
there are no detailed investigation performed on the He
incorporation and diffusion behavior in ZrO$_{2}$ through a
state-of-the-art first-principles approach. To make a deep
understanding of the effect of He in ZrO$_{2}$ in atomic scale is
helpful for us to be more confident to validate the application of
ZrO$_{2}$ in severe environment. Therefore, this is the main purpose
of the present work. In the present study, first the behavior of He
in cubic ZrO$_{2}$ is investigated by determining the formation
energy of point defects. Then the incorporation energy of He into
different sites of the lattice is calculated, through which the most
favorable incorporation site is determined. By taking thermodynamic
equilibrium into account, the solubility of He is also evaluated.
After exploration of the stability of He, we subsequently
concentrate on the diffusion mechanism of He in ZrO$_{2}$ by
calculating energy barriers and migration pathways between two
trapping sites. After comparison of calculated results for different
postulated mechanisms, we present the viable diffusion mechanism of
He in ZrO$_{2}$ matrix.

\section{Computational method}

The density functional theory (DFT) total energy calculations are
self-consistently performed in the framework of the frozen core
all-electron PAW method \cite{Blochl} as implemented in the Vienna
ab initio simulation package (VASP) \cite{Kresse}. The electron
exchange and correlation potentials are evaluated within the GGA
introduced by Perdew, Burke, and Ernzerhof (PBE) \cite{PBE}. Twelve
electrons (4s$^{2}$4p$^{6}$4d$^{2}$5s$^{2}$) for zirconium (Zr) and
six electrons (2s$^{2}$2p$^{4}$) for oxygen (O) are taken into
account as valence electrons. To get accurate results, the cutoff
energy of the plane wave expansion is set up to a high value of 800
eV, which was tested to be well converged with respect to the total
energy. The integration over the Brillouin Zone (BZ) is performed
with a grid of special
\emph{k} point-mesh determined according to the Monkhorst-Pack scheme \cite%
{Monkhorst}. After convergence test, 8$\times$8$\times$8 \emph{k}
point-mesh is chosen to calculate the bulk properties with the total
energy difference less than 1 meV per unit cell. In the present
work, large supercells containing up to 96 atoms have been employed
to reduce any artificial error due to the use of a smaller supercell
as it will be shown later. Correspondingly, a 2$\times$2$\times$2
\emph{k}-point grid is used to sample the BZ of the 96-atom
supercell for the modeling of the point defects and He impurity in
ZrO$_{2}$. Relaxation procedures at ground state are carried out
according to the Gaussian broadening with a smearing width of 0.1
eV. For all defect structures, the atomic relaxation is considered
to be completed when the Hellmann-Feynman forces on all atoms are
less than 0.01 eV/\AA . And the volume relaxation of the supercell
is also calculated.

\section{Results and discussions}

\subsection{Crystallography and bulk properties}

\begin{figure}[tbp]
\includegraphics[width=6cm,keepaspectratio]{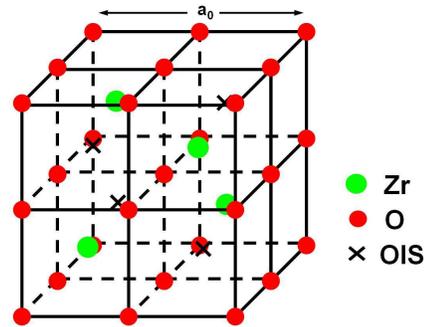}
\caption{Conventional unit cell of ZrO$_{2}$. O atoms are located at
the corners of small cubes, and Zr atoms are located at the center
of an alternative cubes. $\times$ indicates the octahedral
interstitial site (OIS) in the lattice.} \label{fig1}
\end{figure}

Here, we considered the cubic phase of ZrO$_{2}$, which is
crystallized in the fluorite structure with space group
\emph{Fm$\bar{3}$m} (No. 225). The Zr atoms are in the center of a
cube of O atoms, while the O atoms occupy the tetrahedral sites
coordinated by Zr atoms, constituting a simple cubic sublattice as
shown in Fig. 1. The present optimized equilibrium lattice parameter
(\emph{a}) obtained by fitting the energy-volume data in the
third-order Birch-Murnaghan equation of states (EOS) \cite{Birch} is
5.152 \AA\ and some other basic bulk properties calculated with this
PAW method is also listed in Table 1, in fair agreement with the
experimental and previous theoretical results.

\begin{table*}[tbp]
\caption{Calculated equilibrium lattice parameter \emph{a$_{0}$ } (\AA ),
bulk modulus \emph{B$_{0}$} (GPa), and cohesive energy $E_\mathrm{coh}$ (eV)
of fluorite ZrO$_{2}$ at ground state. For comparison, available
experimental values and other theoretical results are also listed.}
\label{bulk}%
\begin{tabular}{lccccccc}
\hline\hline
Parameter & Present work & Previous calculation & Experiment &  &  &  &  \\
\hline
\emph{a$_{0}$ }(\AA ) & 5.152 & 5.130$^{a}$, 5.128$^{b}$, 5.116$^{c}$ & 5.086%
$^{d}$, 5.108$^{e}$, 5.090$^{f}$ &  &  &  &  \\
\emph{B$_{0}$} (GPa) & 234 & 267$^{a}$, 251$^{b}$, 235$^{c}$ & 194$^{g}$, 215%
$^{h}$ &  &  &  &  \\
$E_\mathrm{coh}$ (eV) & -15.99 & -23.28$^{i}$, -11.45$^{j}$ & -11.45$^{k}$,
-14.72$^{l}$ &  &  &  &  \\ \hline\hline
$^{a}$Ref. \cite{Lowther}, $^{b}$Ref. \cite{Jaffe},$^{c}$Ref. \cite{Fadda2},
$^{d}$Ref. \cite{Howard}, &  &  &  &  &  &  &  \\
$^{e}$Ref. \cite{Igawa}, $^{f}$Ref. \cite{Stefanovich}, $^{g}$Ref. \cite%
{Kandil}, $^{h}$ Ref. \cite{Ingel}, &  &  &  &  &  &  &  \\
$^{i}$ Ref. \cite{Jomard}, $^{j}$ Ref. \cite{Medvedeva}, $^{k}$Ref. \cite%
{Samsonov}, $^{l}$ Ref. \cite{Pettifor} &  &  &  &  &  &  &  \\
&  &  &  &  &  &  &
\end{tabular}%
\\[2pt]
\end{table*}

\subsection{Intrinsic point defects formation}

The previous study of ZrO$_{2}$ bulk properties gives insights into the
reliability of the DFT-GGA approach to model the cubic ZrO$_{2}$. Now we
turn to the study of the stability of different types of point defects in
the fluorite lattice. The first-principles method allows us to consider ZrO$%
_{2}$ supercells containing up to hundred atoms and thus to carefully check
the convergence of the defect formation energies as a function of the
supercell size. Indeed, in too small a supercell and due to the periodic
boundary conditions, a point defect interacts with its image in the adjacent
supercell, which yields a spurious contribution to the calculated formation
energy. To check this convergence, we have considered supercells
representing repetition of the elementary cubic cell according to the
following repetition pattern: 1$\times$1$\times$1 (12 atoms), 2$\times$1$%
\times$1 (24 atoms), 2$\times$2$\times$1 (48 atoms), 2$\times$2$\times$2 (96
atoms), 3$\times$2$\times$2 (144 atoms), 3$\times$3$\times$2 (216 atoms), 3$%
\times$3$\times$3 (324 atoms). The defects are explicitly investigated in
the calculation of formation energy as follows: vacancies, interstitials at
OIS, Frenkel pair and Schottky defect.

The formation energies \emph{E$_{F}^{V_\mathrm{X}}$} and \emph{E$_{F}^{I_%
\mathrm{X}}$} of a vacancy (\emph{V$_\mathrm{X}$}) and an interstitial (%
\emph{I$_\mathrm{X}$}) of the \emph{X} specie are obtained from the total
energies of the system with and without the defect, according to
\begin{align}
E_{F}^{V_\mathrm{X}}=E_{}^{V_\mathrm{ZrO_{2}}}-E_\mathrm{ZrO_{2}}+E_\mathrm{X%
},
\end{align}
and
\begin{align}
E_{F}^{I_\mathrm{X}}=E_\mathrm{ZrO_{2}}^{I_\mathrm{X}}-E_\mathrm{ZrO_{2}}-E_%
\mathrm{X},
\end{align}
where \emph{E$_\mathrm{ZrO_{2}}$} is the energy of the defect-free ZrO$_{2}$
supercell, \emph{E$_\mathrm{ZrO_{2}}^{V_\mathrm{X}}$} and \emph{E$_\mathrm{%
ZrO_{2}}^{I_\mathrm{X}}$} are the energies of the supercell
containing one vacancy and interstitial,respectively.
\emph{E$_\mathrm{X}$} is the energy per atom of each chemical
species in its reference state (\emph{X}= Zr or O). Here the
reference state of Zr is chosen as the ground state crystalline
phases, namely, the $\alpha$-Zr crystal with space group
\emph{P6$_{3}$/mmc} (No. 194). As for the O, the atomic oxygen
energy is employed as a reference state \cite{Foster}. In the total
energy calculation of an O atom, an orthorhombic supercell with
lattice constants exceeding 15 \AA~ is employed and the
spin-polarization is included. The calculated formation energy of
the O vacancy in cubic ZrO$_{2}$ is in good agreement with previous
theoretical and experimental results \cite{Bredow}.

Frenkel pair consists of a non-interacting vacancy and an interstitial of
the same chemical element. Thus the formation energy can be determined from
the formation energies of the vacancy and of the interstitial calculated
separately, given by
\begin{align}
E_{F}^{FP_\mathrm{X}}=E_{F}^{V_\mathrm{X}}+E_{F}^{I_\mathrm{X}}.
\end{align}
A Schottky defect is a more complex defect consisting of a Zr
vacancy and two O vacancies, all of which are again supposed to be
non-interacting. The formation energy is calculated by
\begin{align}
E_{F}^{Sh}=E_\mathrm{ZrO_{2}}^{V_\mathrm{Zr}}+2E_\mathrm{ZrO_{2}}^{V_\mathrm{%
O}}-3\frac{N-1}{N}E_\mathrm{ZrO_{2}},
\end{align}
where $N$ is the total atom number in the used supercell.

\begin{figure}[tbp]
\includegraphics[width=6cm,keepaspectratio]{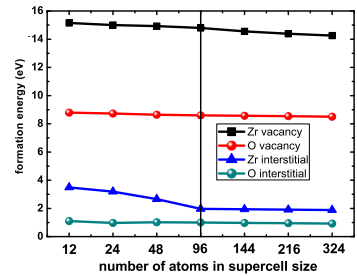}
\caption{Formation energies of single point defects in ZrO$_{2}$ as
a function of the size of the supercell used in the calculation.}
\label{fig2}
\end{figure}

The variation in the calculated formation energies of the single
point defects as a function of the size of the supercell is
represented in Fig. 2. On the whole, clearly, one can see that the
convergence becomes satisfied when the supercell size is chosen as
large as 2$\times$2$\times$2 (96 atoms). Therefore, all the results
reported in the rest of the paper are obtained with the
2$\times$2$\times$2 supercells. Table 2 gives the formation energies
of Zr and O vacancies, interstitials at OIS, Frenkel pair, and
Schottky defect, all of which are obtained after performing the atom
relaxation, and with or
without volume relaxation. The variation in the lattice parameter $\Delta$%
\emph{a}/\emph{a} of the supercell is also reported. This variation is only
indicative of the degree of convergence of the size of the supercell:
ideally, for an infinite-size supercell, this variation should be zero.

\begin{table}[tbp]
\caption{Calculated formation energies \emph{E$_{F}$} (eV) of point defects
in ZrO$_{2}$: Zr and O vacancies (Vac), octahedral interstitials (Int),
Frenkel pair (FP), and Schottky defect (Sh). The energies are calculated
with atom relaxation (pos) and with atom and volume relaxation (pos+vol). $%
\Delta$\emph{a}/\emph{a} is the variation in the lattice parameter of the
96-atom supercell.}
\label{formation energy}%
\begin{tabular}{lcccccccccccccc}
\hline\hline Parameters & Int Zr & Int O & Vac Zr & Vac O & FP Zr &
FP O & Sh &  &  &  & &  &  &  \\ \hline \emph{E$_{F}$} (eV) pos &
2.14 & 1.19 & 15.27 & 8.60 & 17.40 & 9.79 & 18.68
&  &  &  &  &  &  &  \\
\emph{E$_{F}$} (eV) pos+vol & 1.97 & 1.17 & 15.26 & 8.60 & 17.22 &
9.77 &
18.66 &  &  &  &  &  &  &  \\
$\Delta$\emph{a}/\emph{a} & 0.5\% & 0.1\% & 0.1\% & -0.1\% & - & - & 0.2\% &
&  &  &  &  &  &  \\ \hline\hline
\end{tabular}%
\\[2pt]
\end{table}

It can be seen from Table 2 that the formation energies of various
point defects are positively large, consistent with the high
resistance property of ZrO$_{2}$ against radiation damage.
Furthermore, Table 2 shows that the O defects have much lower
formation energies than Zr defects. Thus radiation damage and
deviation from stoichiometry will be preferably accommodated in the
O sublattice: by the formation of O vacancies in hypostoichiometric
ZrO$_{2}$ and by the formation of O interstitials in
hyperstoichiometric ZrO$_{2}$, for instance. The effect of volume
relaxation on the formation energies is negligible, except for the
Zr interstitial for which it is 0.2 eV. For this defect, the
supercell size variation is the largest, but does not exceed 1\%.
Unfortunately, to our knowledge, the experimental values of the
formation energies of point defects for cubic ZrO$_{2}$ are not
available in the literature for comparison.

\subsection{Stability of He impurity in ZrO$_{2}$}

The stability of He at pre-existing trap sites is determined by calculating
the incorporation energy which is defined as the energy needed to trap a He
at a pre-existing trap site (vacancy or interstitial considered here) in the
expression as follows:
\begin{align}
E_\mathrm{inc}=E_\mathrm{ZrO_{2}}^\mathrm{He}-E_\mathrm{ZrO_{2}}-E_\mathrm{He%
},
\end{align}
where $E_\mathrm{ZrO_{2}}^\mathrm{He}$ is the energy of ZrO$_{2}$
supercell containing the He impurity, $E_\mathrm{ZrO_{2}}$ is the
energy of the supercell with an empty trap site of He, and
$E_\mathrm{He}$ is the energy of an isolated He. Therefore, a
positive value means that the energy is required to incorporate He
in the lattice whereas a negative result implies that incorporation
is energetically favorable. A comparison of incorporation energies
is the most simple way to assess the stability of the impurity and
predict the most stable solution site for He provided that trap
sites are available for occupation \cite{Donnelly}. This will be the
situation when the concentration of He atoms is low enough that
incorporation proceeds through occupation of intrinsic defect sites.

The use of incorporation energy is limited since it is not sensitive to any
equilibrium between trap sites. To take into account the equilibrium between
the different trap sites one should consider solution energies. Solution
energies should be used to express the population of He in the different
sites when complete thermodynamical equilibrium is achieved. For a certain
site \emph{X}, the solution energy of He is defined as the incorporation
energy plus the apparent formation energy of the trap site. It is clear that
for an interstitial site incorporation and solution energies are equal. For
vacancy insertion sites, the expression is given \cite{Crocombette} as
\begin{align}
E_\mathrm{sol}=E_\mathrm{inc}+E_{F_\mathrm{app}}^{V_\mathrm{X}},
\end{align}
where E$_{F_\mathrm{app}}^{V_\mathrm{X}}$ is the apparent formation energy
of the vacancy site. With stoichiometry at the ground state, the apparent
formation energies of Zr and O vacancy sites can be expressed \cite%
{Crocombette} as
\begin{align}
E_{F_\mathrm{app}}^{V_\mathrm{Zr}}=E_{F}^{Sh}-E_{F}^{FP_\mathrm{O}}
\end{align}
and
\begin{align}
E_{F_\mathrm{app}}^{V_\mathrm{O}}=\frac{1}{2}E_{F}^{FP_\mathrm{O}}.
\end{align}
Here $E_{F}^{FP_\mathrm{O}}$ and $E_{F}^{Sh}$ are the O Frenkel pair
and Schottky defect formation energies.

We investigate the He behavior at various locations including the
high-symmetry positions in the lattice, and find that He always
moves to the OIS during atomic relaxation in a defect-free
ZrO$_{2}$. For defective structures, we take considerations of trap
sites with one Zr or O vacancy as the substitution site,
respectively. The incorporation energies and solution energies are
calculated by taking into account the relaxation of the atomic
positions in the supercell, with and without the volume relaxation,
and for the different incorporation sites. In addition, through
Bader analysis \cite{Bader,Tang}, we have also calculated the
electron charge density of He when it occupies various trap sites.
All the relevant results are presented in Table 3.

\begin{table}[tbp]
\caption{Incorporation \emph{E$_\mathrm{inc}$} (eV) and solution energies
\emph{E$_\mathrm{sol}$} (eV) of He in ZrO$_{2}$ at different sites of the
fluorite lattice: the OIS, the Zr and the O substitution sites with
relaxation of the atom position (pos) and volume of the supercell (vol),
change in the supercell lattice parameter $\Delta$\emph{a}/\emph{a}, and the
electron charge density \emph{Q} of He at the trap sites.}
\label{Incorporation energy}%
\begin{tabular}{lcccccccccccccc}
\hline\hline
Parameters & He(OIS) & He(Zr) & He(O) &  &  &  &  &  &  &  &  &  &  &  \\
\hline
\emph{E$_\mathrm{inc}$} (eV) pos & 1.3491 & 0.4332 & 1.5931 &  &  &  &  &  &
&  &  &  &  &  \\
\emph{E$_\mathrm{inc}$} (eV) pos+vol & 1.3457 & 0.4199 & 1.5779 &  &  &  &
&  &  &  &  &  &  &  \\
\emph{E$_\mathrm{sol}$} (eV) pos & 1.3491 & 7.1327 & 6.4871 &  &  &  &  &  &
&  &  &  &  &  \\
\emph{E$_\mathrm{sol}$} (eV) pos+vol & 1.3457 & 7.1176 & 6.4650 &  &  &  &
&  &  &  &  &  &  &  \\
$\Delta$\emph{a}/\emph{a} (\%) & 0.03 & 0.15 & 0.12 &  &  &  &  &  &  &  &
&  &  &  \\
\emph{Q} (electrons) & 2.0596 & 2.0298 & 2.1891 &  &  &  &  &  &  &  &  &  &
&  \\ \hline\hline
\end{tabular}%
\\[2pt]
\end{table}

The calculated incorporation energies at any trap site are positive
indicating that they are less stable in ZrO$_{2}$ fluorite lattice
than in the free gas phase. This means that some energy has to be
provided to incorporate He in ZrO$_{2}$. The minimum incorporation
energy is found for the Zr vacancy. With relatively high
incorporation energies of He at OIS and O vacancy, we expect a
slight preference for He to be located at Zr vacancy. However, the
large solution energies at
various sites (see Table 3) lead to the prediction of the insolubility of He in cubic ZrO$%
_{2}$. We can conclude that the driving force for the experimentally
observed He release has its origin in this insolubility. The results
shows that He is more likely to be accommodated at the OIS for the
thermodynamic equilibrium concentration. As for the perturbation of
the crystal induced by the impurity, it is revealed that the volume
relaxation with the incorporated He has no great influence on the
incorporation energies and the He impurity only induces a slight
expansion of the lattice parameter. Finally, Bader analysis results
show that the He impurity displays only a weak charge transfer, in
agreement with a complete electronic shell and neutral environment
at this position. This finding also explains why a Zr vacancy is the
most favorable incorporation site for He since it is well known
\cite{Puska} that a He atom prefers to occupy the site with the low
electron density due to its filled-shell electronic configuration.

\subsection{Diffusion of He impurity in ZrO$_{2}$}

The concentration of He remaining in the lattice is decided not only
by the solution energies but also by the ease with which the He atom
can diffuse out of the lattice. Here, we take considerations into
the possible diffusion mechanisms for one He atom. The climbing
image nudged elastic band (CINEB) method \cite{Henkelman} is
employed to find the minimum energy path (MEP) of atomic migration
when both the initial and final configurations are known. The CINEB
is an efficient and improved method for finding the energy saddle
point between the given initial and final states of a transition to
understand the energy path of He diffusion behavior. Migration
energy barrier is defined as the energy difference between an
initial configuration and the saddle point. The interstitial and
vacancy assisted diffusion mechanisms of one He atom in ZrO$_{2}$
are systematically studied as follows.

\subsubsection{interstitial diffusion mechanism}

We first investigate the interstitial diffusion mechanism of He in
defect-free ZrO$_{2}$ by calculating the migration energy between
two adjacent OIS sites, which are the body-centered position of the
lattice. The diffusion pathway and MEP are shown in Fig. 3. We
perform the CINEB calculation in which the initial guess for the
atomic positions of the diffusing He atom is simply a linear
interpolation between the OIS sites. The saddle point obtained with
this initial guess is found to be in the middle of the migration
pathway as shown in Fig. 3(b), corresponding to an energy barrier of
3.74 eV. In the panel of Fig. 3(b), we can see obviously the
distortion of the nearest two O atoms at the saddle point, which are
pushed away from their lattice sites by about 0.37 \AA\ in the
perpendicular direction to the migration pathway of He by the strain
energy.

\begin{figure}[tbp]
\includegraphics[width=6cm,keepaspectratio]{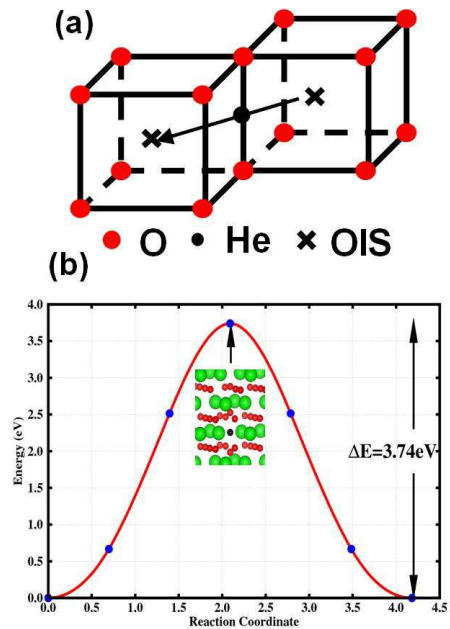}
\caption{(a) Diffusion pathway and (b) MEP of He migration from OIS
into the nearest OIS in defectfree ZrO$_{2}$.} \label{fig3}
\end{figure}

\subsubsection{Zr vacancy assisted mechanism}

Next, we explore the migration pathways of He in a defective ZrO$_{2}$ and
find that the energy barriers are remarkably decreased compared to the
previous results. We take Zr and O vacancy into account separately to probe
into the minor difference of the vacancy assisted diffusion mechanism. In
the case of Zr vacancy preexisting, three possible migration pathways are
assumed and CINEB calculations are carried out similarly.

The first postulated mechanism is involved in the direct diffusion of He
from OIS into the nearest OIS in the presence of Zr vacancy, shown in Fig.
4(a). Examining the MEP for this mechanism in Fig. 4(b), two images on
either side of the saddle point configuration exhibit a slight oscillation
in energies around the transition state. The amplitude of the oscillation is
only 0.02 eV, which is within the error of the calculations. The reliable
estimate of the energy barrier can be obtained as about 0.23 eV, which must be
overcome in the direct diffusion path of He from OIS to OIS. We find that
the Zr vacancy is a trap for He atom from the MEP. Along the diffusion
pathway, He atom moves into the Zr vacancy during the atomic relaxation at
the third image, and an additional energy of about 1.36 eV must be provided when
the He migrates back into the nearest OIS as shown in Fig. 4(b). The
distortion of the nearest O atoms of He is calculated. Due to the strain
energy of He, the displacement of O atom in the perpendicular direction to
the pathway is about 0.43 \AA\ at the saddle point.

\begin{figure}[tbp]
\includegraphics[width=6cm,keepaspectratio]{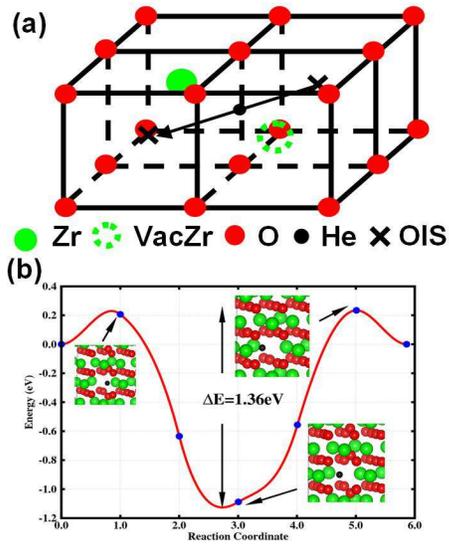}
\caption{(a) Diffusion pathway and (b) MEP of He migration from OIS
into the nearest OIS in the presence of Zr vacancy in ZrO$_{2}$.}
\label{fig4}
\end{figure}

The second kind of diffusion is assumed that He hops through a Zr
vacancy from one OIS to the next nearest OIS. In addition, the two
adjacent OIS lies in the line with Zr vacancy, represented in Fig.
5(a). It is interestingly noted that the MEP in Fig. 5(b) displays
the same migration energy trend as the first one in Fig. 4(b). The
energy barrier is 0.23 eV for migration from OIS to the nearest OIS,
and once the He atom diffuses into the Zr vacancy, it hardly moves
into the next OIS unless an extra energy of 1.36 eV is gained. The
next nearest O atom has the largest distortion 0.48 \AA\ when He is
trapped in the Zr vacancy.

\begin{figure}[tbp]
\includegraphics[width=6cm,keepaspectratio]{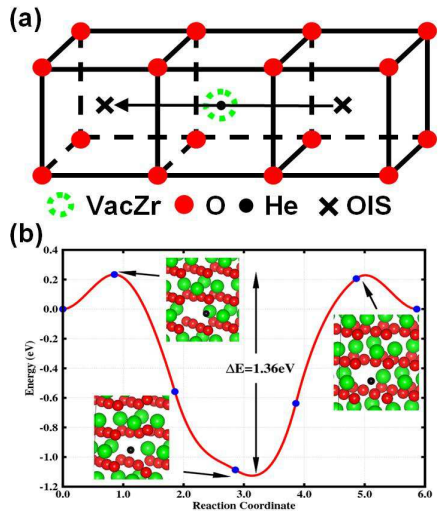}
\caption{(a) Diffusion pathway and (b) MEP of He migration from OIS
into the next nearest OIS in the presence of Zr vacancy in
ZrO$_{2}$.} \label{fig5}
\end{figure}

The third possible diffusion path is similar with the second
assumption, but the two nearest OIS is not in the line with Zr
vacancy, as illustrated in Fig. 6(a). We consider the CINEB
calculation of this migration in two steps. The MEP in Fig. 6(b)
shows the step that the He atom first jumps into the Zr vacancy from
the OIS, and Fig. 6(c) gives the situation that He atom migrates
into the nearest OIS from Zr vacancy. From initial OIS to Zr
vacancy, the saddle point is obtained with an energy barrier of 0.23
eV, and when He diffuses from Zr vacancy to final OIS, another
saddle point appears with a larger migration barrier of 0.91 eV.
Similarly, in this case the distortion of the nearest O atoms of He
at Zr vacancy is as small as only 0.10 \AA .

From above analysis, it is concluded that in ZrO$_{2}$, as Zr
vacancy preexists, He diffuses easily with a lower migration energy
barrier than the interstitial diffusion mechanism. Different
diffusion pathways proposed here demonstrate a similar migration
characteristic. Obviously, it is more likely to trap the He atom
into the Zr vacancy along the diffusion pathway when compared to the
defect-free cases.

\begin{figure}[tbp]
\includegraphics[width=6cm,keepaspectratio]{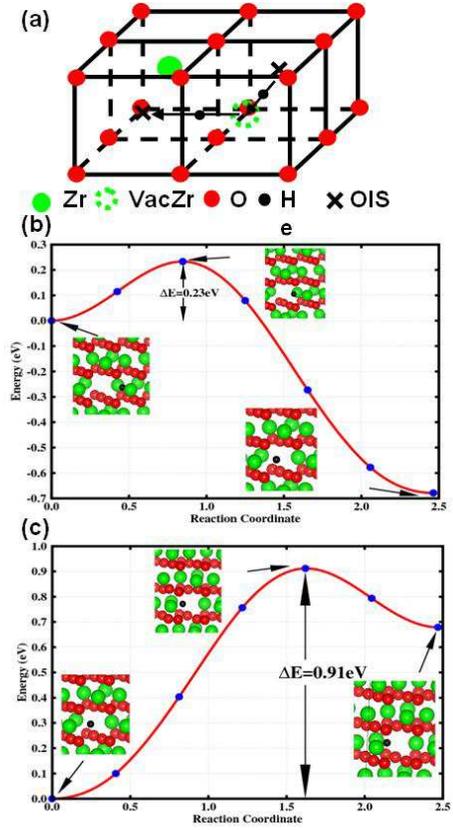}
\caption{(a) Diffusion pathway, (b) MEP of He migration from OIS
into Zr vacancy, and (c) MEP of He migration from the Zr vacancy
into the nearest OIS.} \label{fig6}
\end{figure}

\subsubsection{O vacancy assisted mechanism}

As for O vacancy that preexists in ZrO$_{2}$, similar consideration
is taken to evaluate the diffusion mechanism of He atom. At first,
we discuss that He atom directly migrates from OIS into the nearest
OIS, given in Fig. 7(a). Then He atom hopping from OIS to nearest
OIS through O vacancy is investigated. It is considered that the He
atom first moves into the O vacancy from the OIS, and next the He
atom is assumed to jump into the nearest OIS from the O vacancy,
shown in Fig. 8(a). The MEP of direct migration from OIS to the
nearest OIS in the presence of O vacancy is different from the
migration with the preexisting Zr vacancy discussed above, as shown
in Fig. 7(b). There are two images in Fig. 7(b) that are predicted
to be most energetically favorable. In the middle point of the
pathway, the metastable state is found between the two local saddle
points. At the middle image of this pathway, the He atom is trapped
into the O vacancy after the relaxation during CINEB calculation.
The energy barrier from the energy favorable site to the saddle
point is 2.18 eV. On the other hand, the MEP of He atom hopping from
OIS to nearest OIS through O vacancy is illustrated separately in
Fig. 8(b) and Fig. 8(c). The saddle point is obtained in the pathway
with the energy barrier 1.04 eV and 0.92 eV, corresponding the
migration energy barrier from the initial OIS to O vacancy and from
O vacancy to the final OIS, respectively. At various images in the
CINEB calculation, the O atoms near the O vacancy move inward to the
vacancy at about 0.25\AA .

\begin{figure}[tbp]
\includegraphics[width=6cm,keepaspectratio]{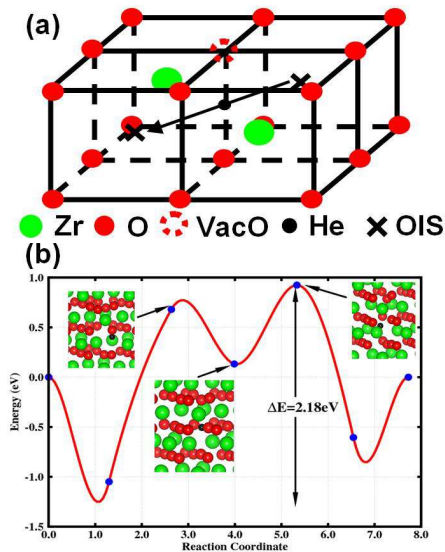}
\caption{(a) Diffusion pathway and (b) MEP of He migration from OIS
into the nearest OIS in the presence of O vacancy in ZrO$_{2}$.}
\label{fig7}
\end{figure}

\begin{figure}[tbp]
\includegraphics[width=6cm,keepaspectratio]{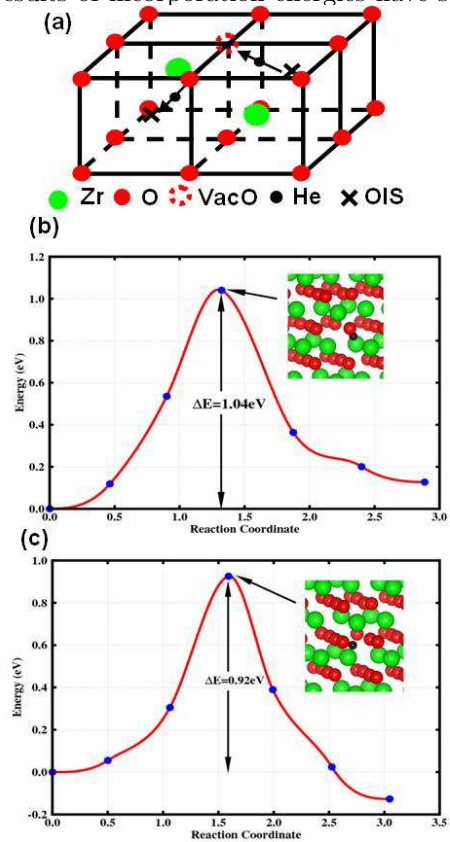}
\caption{(a) Diffusion pathway, (b) MEP of He migration from OIS
into O vacancy, and (c) MEP of He migration from the O vacancy into
the nearest OIS. } \label{fig8}
\end{figure}

From the above discussion, we find that the migration barrier of
interstitial diffusion is relatively higher than the vacancy
assisted diffusion. During the migration of He with Zr or O vacancy
preexisted, the He atom is easily trapped into the vacancy and an
extra energy of only about 1 eV is needed to hop out of the vacancy,
which remarkably agrees well with the experimental result
\cite{Damen}. Therefore, the diffusion behavior of He atom in cubic
ZrO$_{2}$ is dominated by the vacancy assisted diffusion mechanisms.

\section{CONCLUSION}

In summary, we have performed total-energy calculations to
investigate the formation energies of native point defects, the He
incorporation and solution energies, and the diffusion properties of
He in cubic ZrO$_{2}$, using the PAW-GGA method and supercell
approach. Our calculated results of incorporation energies have
suggested that trapping of a He atom at a Zr vacancy is more stable
than at a O vacancy or at an OIS. From the solution energy results,
on the other hand, it has been found that the He is more likely to
be located at the OIS for the thermodynamic equilibrium
concentration. To get more insights into the diffusion mechanism of
He, we have calculated the migration energy of He in various
pathways through the CINEB calculation, from which it has been shown
once the He is trapped by the vacancy, its migration energy barrier
from the vacancy to the nearest OIS is largely decreased when
compared to the vacancy-free case. Thus, we conclude the diffusion
of He in ZrO$_{2}$ is mainly assisted by the vacancy, which is
expected to provide a guiding line in explaining the experimentally
observed He diffusion phenomenon.

\section{ACKNOWLEDGMENTS}

This work was supported by NSFC under Grant No. 51071032 and by the
National Basic Security Research Program of China.

\end{document}